\documentclass[reprint,aps,pra,superscriptaddress,floatfix]{revtex4-2}
\usepackage{blindtext}
\usepackage{parskip}
\usepackage{xcolor}
\usepackage{mathrsfs}
\usepackage{siunitx} % or simply amstext
\usepackage{graphicx}% Include figure files
\usepackage{dcolumn}% Align table columns on decimal point
\usepackage{bm}% bold math
\usepackage{amsmath}
\usepackage[utf8]{inputenc}
\usepackage[T1]{fontenc}
\usepackage{mathptmx}
\usepackage{etoolbox}
\usepackage[update,prepend]{epstopdf}
\usepackage{pifont}% For special characters
\usepackage[breaklinks=true,colorlinks=true,linkcolor=blue,urlcolor=blue,citecolor=blue]{hyperref}
\usepackage{tikz,xcolor,hyperref}
\hypersetup{colorlinks=true,linkcolor=blue,urlcolor=blue,citecolor=blue,pdfstartview={XYZ null null 1.00},pdfauthor={Subhendu Kahaly}}
\definecolor{lime}{HTML}{A6CE39}
\DeclareRobustCommand{\orcidicon}{%
	\begin{tikzpicture}
	\draw[lime, fill=lime] (0,0) 
	circle [radius=0.16] 
	node[white] {{\fontfamily{qag}\selectfont \tiny ID}};
	\draw[white, fill=white] (-0.0625,0.095) 
	circle [radius=0.007];
	\end{tikzpicture}
	\hspace{-2mm}
}

\foreach \x in {A, ..., Z}{%
	\expandafter\xdef\csname orcid\x\endcsname{\noexpand\href{https://orcid.org/\csname orcidauthor\x\endcsname}{\noexpand\orcidicon}}
}

\begin{document}
%\preprint{}

% \title{Time Resolved Characterization of High Repetition Rate Gas Jet Target For High Harmonic Generation}

% \title{Temporal evolution of High Repetition Rate Gas Jet Target For High Harmonic Generation}

\title{Time Resolved Investigation of High Repetition Rate Gas Jet Target For High Harmonic Generation}

\author{Balázs Nagyillés}
\email[Correspondence:~]{balazs.nagyilles@eli-alps.hu}
\affiliation{ELI ALPS, ELI-HU Non-Profit Ltd., Wolfgang Sandner utca 3., Szeged 6728, Hungary}%
\affiliation{Institute of Physics, University of Szeged, D\'om t\'er 9, H-6720 Szeged, Hungary}%Lines break automatically or can be forced with \\
\author{Zsolt Diveki}%
\affiliation{ELI ALPS, ELI-HU Non-Profit Ltd., Wolfgang Sandner utca 3., Szeged 6728, Hungary}%
\author{Arjun Nayak}%
\affiliation{ELI ALPS, ELI-HU Non-Profit Ltd., Wolfgang Sandner utca 3., Szeged 6728, Hungary}%
\author{Mathieu Dumergue}
\affiliation{ELI ALPS, ELI-HU Non-Profit Ltd., Wolfgang Sandner utca 3., Szeged 6728, Hungary}%
\affiliation{LULI–CNRS, CEA, Sorbonne Université, Ecole Polytechnique, Institut Polytechnique de Paris, Paris}
% \author{Sergei Kühn}%
% \affiliation{ELI-ALPS, ELI-HU Non-Profit Ltd., Wolfgang Sandner utca 3., Szeged 6728, Hungary}%
\author{Bal\'{a}zs Major}
\affiliation{ELI ALPS, ELI-HU Non-Profit Ltd., Wolfgang Sandner utca 3., Szeged 6728, Hungary}%
\affiliation{Department of Optics and Quantum Electronics, University of Szeged, D\'om t\'er 9, H-6720 Szeged, Hungary}
\author{Katalin Varj\'{u}}
\affiliation{ELI ALPS, ELI-HU Non-Profit Ltd., Wolfgang Sandner utca 3., Szeged 6728, Hungary}%
\affiliation{Department of Optics and Quantum Electronics, University of Szeged, D\'om t\'er 9, H-6720 Szeged, Hungary}
\author{Subhendu Kahaly\orcidA{}}%
\email[Correspondence:~]{subhendu.kahaly@eli-alps.hu}
\affiliation{ELI ALPS, ELI-HU Non-Profit Ltd., Wolfgang Sandner utca 3., Szeged 6728, Hungary}%
\affiliation{Institute of Physics, University of Szeged, D\'om t\'er 9, H-6720 Szeged, Hungary}
%%%%%%%%%%%%%%%%%%%%%%%%%%%%%%%%%%%%%%%%%%%%%%%%%%%%%%%%%%%%%%%%%%%%%%%%%%%%%%%%%%%%%%%%%%%%%%%%%%%%%%%%%

\date{\today}

\begin{abstract}
High repetition rate gas targets constitute an essential component in intense laser matter interaction studies. The technology becomes challenging as the repetition rate approaches kHz regime. In this regime, cantilever based gas valves are employed, which can open and close in tens of microseconds, resulting in a unique kind of gas characteristics in both spatial and temporal domain. Here we characterize piezo cantilever based kHz pulsed gas valves in the low density regime, where it provides sufficient peak gas density for High Harmonic Generation while releasing significantly less amount of gas reducing the vacuum load within the interaction chamber, suitable for high vacuum applications. In order to obtain reliable information of the gas density in the target jet space-time resolved characterization is performed. The gas jet system is validated by conducting interferometric gas density estimations and high harmonic generation measurements at the Extreme Light Infrastructure Attosecond Light Pulse Source (ELI ALPS) facility. Our results demonstrate that while employing such targets for optimal high harmonic generation, the high intensity interaction should be confined to a suitable time window, after the cantilever opening. The measured gas density evolution correlates well with the integrated high harmonic flux and state of the art 3D simulation results, establishing the importance of such metrology. 
\end{abstract}

% \pacs{}% insert suggested PACS numbers in braces on next line

\maketitle %\maketitle must follow title, authors, abstract and \pacs

\section{\label{sec:introduction}Introduction}

Investigations in ultrashort laser-plasma science in the strong field regime are generically based on the interaction of an appropriately focused laser driver on to reflective (overdense) or transparent (underdense) targets. The interaction conditions needs to be reproduced, and hence the target needs to be replenished, at the repetition rate of the laser. Recent advances in few cycle, high peak power high repetition rate ($\geq$ 1 kHz) lasers \cite{toth_sylos_2020,Ouill2020,Stanfield2021,Furch2022} has expedited the development and characterization of targets that are able to sustain interactions at such challenging repetition rate in a reproducible and stable manner. For all transmission based experiments in this domain, the use of gas targets is widespread because they can provide dense, stable and reproducible medium for laser matter interaction studies. 

The application space is ever expanding with recent demonstrations of laser wake field acceleration of electrons \cite{Gunot2017,Huijts2022} and high harmonic based \emph{attosecond} pulse generation \cite{Mikaelsson2020,Witting2022,Ye2022,Csizmadia2023}, both operating at a high repetition rate. In both the cases a continuous gas cell has been used for the interaction and the accessible gas density space is limited due to the residual gas load within the vacuum chamber. One straight-forward way to overcome this is to use a high repetition rate gas jet target with appropriate nozzle geometry. Pulse valves working up to a very high pressure and gas density has been demonstrated \cite{Sylla2012}, albeit operating at a low frequency. Nonetheless the available repetition rate for pulsed valves currently allows one to reach up to $\sim$ 5kHz \cite{Irimia2009, Irimia2009ppcf,Even2014, Meng2015}. The importance of careful metrology of gas jets emanating from such valves with respect to their appropriate application space cannot be overemphasized. Such systems are important for the \emph{attoscience} community and beyond. For example coupled with the emergence of $\geq$ 1 kHz intense lasers \cite{toth_sylos_2020,Ouill2020,Stanfield2021,Furch2022} such a high repetition rate gas jet target can enable the extension of the recent demonstrations like multi millijoule THz \cite{Pak2023} and/or relativistic single cycle mid IR pulses \cite{Zhu2020, Nie2020} to the high average power regime, opening up wide ranging applications. The capability of solenoid type Even-Lavie valves operating at less than 2 kHz repetition rate has been demonstrated in the domain of high harmonic spectroscopy of molecules \cite{Ren2013,Tro2017} and transient absorption spectroscopy \cite{Leshchenko2023}. Here we undertake the space and time resolved investigation of the gas density profile of a piezo cantilever based high repetition rate gas jet from the perspective of optimizing the high harmonic generation (HHG). 
%%%%%%%%%%%%%%%%%%%%%%%%%%%%%%%%%%%%%%%%%%%%%%%%%%%%%%%%%%%%%%%%%%%%%%%%%%%%%%%%%%%%%%%%%%%%%%%%%%%%%%%%%%%%%%%%
\begin{figure*}[ht!]\label{fig:summary_pc}
    \includegraphics[width=0.99\textwidth]{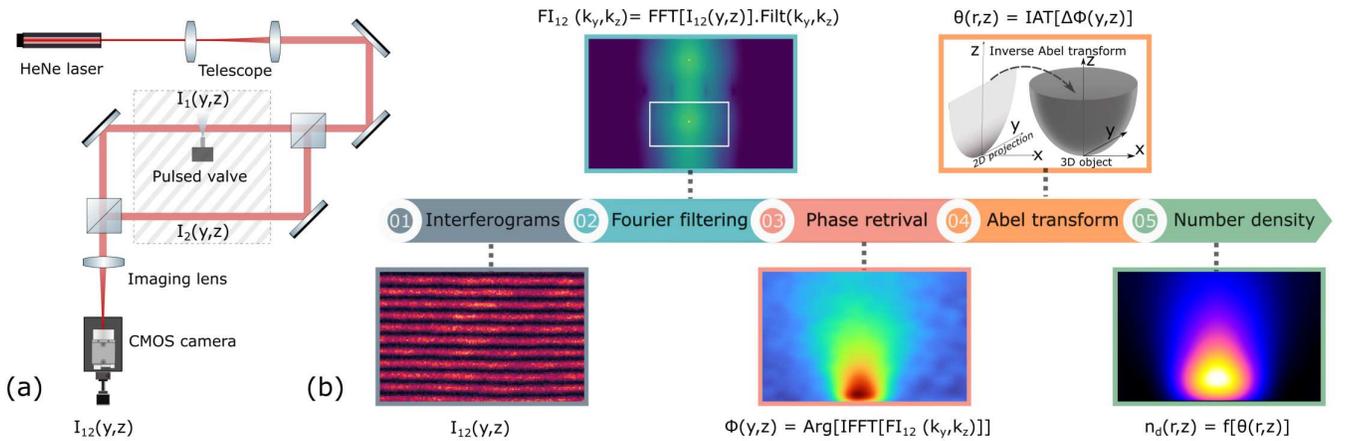}%
    \caption{\label{fig:gasnogas}Interferometric setup for gas density profile measurements (a) for the gas jet based on high frequency cantilever valve and data analysis flow (b) for number density extraction from the measured interferograms. (a) The schematic layout of the pulsed jet density characterisation setup. A Mach-Zehnder setup is used to extract the projection interferograms $I_{12}(y,z)$. The camera images the gas jet plane and $I_{12}(y,z)$ is formed by interference of the spatial field distribution in the gas jet plane (interaction arm %$1$
    indicated by $I_{1}(y,z)$) and an equivalent plane in the other arm (reference arm %$2$
    indicated by $I_{2}(y,z)$). (b) Steps of the Abel transform based gas density extraction. (01) Acquisition of interference pattern from the interferometer. (02) 2D Fourier transform of the interference pattern and filtering the zone of interest in the Fourier space with an appropriate mask. (03) The accumulated phase $\phi(y,z)$ due to optical path difference between the two arms is extracted from Inverse Fourier transform of $FI_{12}(k_{y},k_{z})$. For each gas density extraction, $\phi(y,z)$ is retrieved %for the case 
    with gas ($\phi_{g}(y,z)$) and without gas ($\phi_{0}(y,z)$). The difference between these two terms give the contribution of phase due to the presence of gas, $\Delta \phi(y,z) = \phi_{g}(y,z)-\phi_{0}(y,z)$. In (04) inverse Abel transform (IAT) is utilised on $\Delta \phi(y,z)$ (on the retrieved path difference), to obtain the radial and vertical phase variation $\theta(r,z)$ in the laser beam due to the gas jet. Finally in (05) this phase variation is converted to the refractive index, $n(r,z) = \frac{\lambda}{2\pi}\theta(r,z)+1$ and the number density $n_{d}(r,z)$ is extracted from the functional dependence of $n$ on $n_{d}$. The images in (b) represents the typical data at each step of the analysis protocol.}
\end{figure*}
%%%%%%%%%%%%%%%%%%%%%%%%%%%%%%%%%%%%%%%%%%%%%%%%%%%%%%%%%%%%%%%%%%%%%%%%%%%%%%%%%%%%%%%%%%%%%%%%%%%%%%%%%%%%%%%%

HHG is a non-linear process where the strong fundamental laser field gets coherently upconverted to a comb of higher frequency radiation \cite{Ferray1988}. This frequency conversion happens in a gas target most of the cases, when the atomic/molecular system of the gas is driven in the strong field regime \cite{Nayak2019,Amini2019}. The conversion efficiency is inherently defined by the HHG process which is dependent on the characteristics of the generating laser pulse and the gas medium. One of the parameters to optimize the high harmonic radiation, is the pressure of the gas target, since the number of particles determine the number of emitters and absorbers in the HHG and define the phase-mismatch. There is a fine balance between increasing the number of emitters and absorbing the generated harmonics \cite{Constant1999,Major2021}, for a given set of laser parameters. Thus, it is evident that proper gas target characterization is essential for optimization of the HHG process.

In this article, we perform interferometric characterization of the space-time resolved gas density profile and study the HHG from a  cantilever based high repetition rate piezo gasjet system. Our investigation reveal a clear correlation of the HHG yield and the dynamics of the gas density evolution. We further corroborate our observation with 3D strong field simulations that incorporate, microscopic HHG along with macroscopic propagation effects emulating the experimental conditions.Our results show that the gas density profile resulting from such a valve is intricately linked to the dynamics of the cantilever piezo. The remarkable correlation of the HHG signal with the gas density dynamics allows us to achieve stable and optimum harmonic signal through careful timing setting of the valve opening with respect to the pulse arrival. This also establishes the importance of such space time resolved characterization for each such high repetition rate piezo cantilever based valve for any given application under consideration. This becomes crucial in systems like the SYLOS COMPACT beamline at ELI-ALPS \cite{Kuhn2017,Charalambidis2017} where several high repetition rate pulsed gas jets can be placed in sequence \cite{Nayak2018}, in order to improve XUV beam energy, by optimising the phase matching conditions for applications in nonlinear XUV physics \cite{Manschwetus2016,Orfanos2020,Orfanos2022}. 

\section{\label{sec:experiments}Experiments}
The required behaviour of a gas jet in HHG is its short opening time, while creating high density jet at its orifice at high repetition rate. It is crucial to reliably synchronize the timing of the nozzle opening of the gas jet and the arrival of the generating laser pulse, in order to get the best harmonic yield. The experiments have been conducted at two separate locations. 
We developed a standalone test station to characterize the gas density inside the jet under different timing of trigger, valve opening time and backing pressure. We used the outcome to compare it to the harmonic yield obtained from the experiments conducted at the SYLOS COMPACT beamline \cite{Kuhn2017,Charalambidis2017} at ELI-ALPS.

\subsection{Gas jet characterization by interferometry}

The density profiling of gas targets has been carried out with several different methods (see \cite{Comby2018} and references teherin), for both static \cite{Comby2018} and pulsed jets \cite{Altucci1996, Brandi2011, Xu2021,Hagmeister2022}. Here we undertake space-time resolved interferometry to access the gas atomic density distribution. The experimental layout is based on an Mach-Zehnder interferometer, Fig.~\ref{fig:gasnogas}(a). An expanded He-Ne laser is used to enter the interferometer after being split in two arms. The interaction arm passes through a gas jet to introduce some phase shift in the optical path of the laser beam with respect to the reference arm. Both arms are in vacuum. The gas target transverse plane (represented by $I_{1}(y,z)$ in Fig.~\ref{fig:gasnogas}(a)) is imaged onto a CMOS sensor (a commercial Basler acA1440-73gm camera), where the recombined beams form the interference pattern ($I_{12}(y,z)$ in Fig.~\ref{fig:gasnogas}(b)). Several vital points have been taken into account when characterizing the gasjets:

\begin{itemize}
    \item To keep the signal to noise ratio high (especially at low gas jet densities) an ATH 500M turbo-molecular pump was providing the low ambient pressure in the chamber, 10$^{-5}$-10$^{-6}$ mbar. Additionally, the whole part of the interferometer, which is not in vacuum had to be covered to protect the beam paths from air fluctuations and the assembly was placed on stable optical table. These precautions, reduced the residual gas load, minimised parasitic vibrations and reduced refractive index fluctuations in the interferometer, allowing the intrinsic noise of the setup to be limited to gas density levels as low as about $3\times10^{17}~cm^{-3}$, estimated from the analysis of reference images of the interference pattern, without activating the valve.
    \item The experimental target gas is argon which has high refractive index of 1.00028 at wavelength $\lambda=$633 nm (for example significantly higher compared to helium 1.000034 at the same $\lambda$) allowing for more sensitivity in terms of measuring phase difference, in spite of the sub-millimetric width of the gasjet, even in the sub-$10^{18}~cm^{-3}$ density regime.
    \item The gas refractive index \emph{ansatz} is valid, when the characterization is not corrupted due to molecular jet formation with large clusters. Within the parameter range relevant to us the empirical Hagena parameter $\Gamma^{*}\ll 100$ is significantly less that the limit $\Gamma^{*}\sim 10^{3}$ required for cluster formation \cite{Hagena1981, Hagena1992,Nam2020}.
\end{itemize}

For the tests we used an Amsterdam Piezo Valve ACPV2 model, a cantilever piezo with 500 $\mu$m nozzle size. Cantilever piezos can deliver large displacements up to hundreds of micrometers to 1 mm, while working at high repetition rates, up to 5kHz. The difference of the cantilever piezos to disk shaped piezos is that by adjusting the free length of the cantilever one can adjust the displacement of the cantilever\cite{Irimia2009}. For example, by decreasing the length, the displacement drops rapidly, while its resonant frequency increases. Cantilever resonance can introduce observable effects in gas density measurement. Since the cantilever will bounce back and forth while opening and closing the pulsed valve, it can introduce pressure and hence number density fluctuation in the released gas within one such cycle of operation.

The synchronization between the camera and the jet is realized with a delay generator. The time resolution of the measurement - which is determined by the shortest possible exposure time of the CMOS sensor is 1 $\mu$s. For each measurement two images are recorded, one with the nozzle opened and one with the nozzle closed serving as a reference measurement without any gas present in any arms - this is realized by running the camera at twice the repetition rate of the gas source. For resolving the temporal evolution of the gas density while opening the valve, the camera trigger was delayed compared to the trigger signal of the jet. The setup can record images at up to 100 Hz, but in order to get a background free image one has to wait until the turbomolecular pump can reduce the pressure in the chamber to the base - 10$^{-5}$-10$^{-6}$ level, resulting a few Hertz operation.

As depicted in the flowchart in Fig.~\ref{fig:gasnogas}(b), the 2D phase shift $\phi(y,z)$ is extracted from the interferogram using 2D Fourier transformation algorithm described in Ref. \cite{Takeda1982}. One can see from a typical unwrapped phase map presented in Figure \ref{fig:gasnogas}(b) (step 3) that in the plane perpendicular to the propagation axis $x$, the jet rapidly spreads out as the distance from the nozzle tip increases (vertical $z$ direction). The extra contribution to the phase shift introduced in the probe beam propagating along $x$ by the argon gas density profile is, $\Delta \phi(y,z) = \int \frac{2\pi}{\lambda}\Delta n(x,y,z)dx$, where $\Delta n(x,y,z)=n(x,y,z)-1$, is the shift in index of refraction due to the presence of the gas and $n(x,y,z)$ is the refractive index of argon jet. As explained in the caption of Fig.~\ref{fig:gasnogas}, $\Delta \phi(y,z)$ is calculated from two projection interferograms: one with the gasjet on and the other without any gas in the interaction arm.

The measured phase-map $\Delta \phi(y,z)$ is a 2D projection of the 3D distribution of the
phase difference $\Delta \phi(r,z)$ introduced by the gasjet ($r$ is the radial distance from the center of the gasjet axis $z$). Since one can assume that the jet has a cylindrical symmetry it is
possible to transform the projection $\Delta \phi(y,z)$ to a radial distribution $\theta(r, z) = \Delta \phi(r,z)$ using the inverse
Abel transform (IAT) \cite{Abel1826} as follows:
\begin{equation}\label{eq:rho}
    \theta(r, z) = IAT[\Delta\phi(y, z)] = -\frac{1}{\pi}\int\limits_r^\infty\frac{d\Delta\phi(y, z)}{dy}\frac{1}{\sqrt{y^2 - r^2}} dy
\end{equation}
% https://arxiv.org/pdf/1902.09007.pdf
where $r$ is the radial distance from the center of the nozzle, $z$ is the vertical distance from the tip of the nozzle and transverse coordinate $y$ is the coordinate perpendicular to $x$ and $z$. This is indicated in step 4 of Fig.~\ref{fig:gasnogas}(b).

We numerically carry out the IAT in Python using the well developed BAsis Set EXpansion (BASEX)\cite{Dribinski2002} method in the package PyAbel \cite{Hickstein2019}.
% As one can see, if $\phi$ is in radians, the dimension of $\rho$ is rad/m.
The refractive index can be expressed with the radial distribution using equation, $\Delta n(r,z)=n(r,z)-1 = \frac{\lambda}{2 \pi} \theta (r,z)$,
where $\lambda$ is the wavelength of the laser. The refractive index is connected to the number density. This can be found through the series of few steps. The molar reflectivity (A) relates the optical properties of the substance with the thermodynamic properties. From the Lorenz—Lorentz  expression which is dependent on the temperature (T) through the molar mass (M), $A = \frac{\left( n^2 -1 \right) }{\left( n^2 +2 \right) } \frac{M}{\rho}$, where $n$ is the refractive index of the atomic gas and $\rho$ is the gas density \cite{Xu2021}. The molar mass can be given by, $M = \frac{R T \rho}{p}$, where $R$ is the universal gas constant and $p$ is pressure. Using the relation between polarizability ($\alpha_e=1.664$ \AA$^{3}$ for argon) and molar reflectivity, $A = \frac{4}{3}N_A\alpha_e\pi$
and ideal gas law as $pV = NRT$, the number density $n_{md}=\frac{N}{V}$ in the units of particles/cm$^3$ can be written as:
\begin{equation}\label{eq:num_density}
   n_d = \frac{n_{md}}{N_A} = \frac{3}{4}\frac{\left( n^2 -1 \right) }{\left( n^2 +2 \right) } \frac{1}{N_A^{2} \alpha_e \pi}
\end{equation}

This is the last step presented in Fig.~\ref{fig:gasnogas}(b).

\subsection{High Harmonic Generation in the beamline}\label{sec:HHG_beamline}

The gas density dynamics during the opening and closing of the cantilever is crosschecked on the SYLOS COMPACT beamline \cite{Kuhn2017,Charalambidis2017}. The main goal of this beamline is to achieve high energy isolated \emph{attosecond} pulses as well as \emph{attosecond} pulse trains in the sub 150 eV regime at high repetition rate, in order to perform XUV-pump XUV-probe nonlinear experiments \cite{Orfanos2020, Orfanos2022}. To achieve this goal it uses long laser focusing (10 m) and up to four high pressure gas jets to generate XUV radiation. The generated XUV beam is separated from the driver IR by a 200 nm thick Al filter and the XUV signal is detected with a calibrated XUV photodiode. The incoming beam size is around 6 cm which was reduced with an iris to 3 cm in order to maximize the XUV yield - resulting around 12 mJ in the interaction region. These conditions enable the generation of around 30$\,$nJ XUV pulses, from argon gas, after XUV filter.

The driving laser for this experiment was the SYLOS Experiment Alignment (SEA) laser \cite{Budriunas2017} which operates at 10$\,$Hz and delivers 34$\,$mJ pulse energy with 11$\,$fs pulse duration at 825$\,$nm central wavelength. When optimizing the XUV energy it is crucial to get the timing of the valve opening and the opening duration correct for the individual gas jets, in order to maximize the gas density in each interaction region. Keeping the valve opening time constant and changing the delay between the laser trigger and the opening of the valve one can study the impact of the dynamics of the valve opening on the integrated yield of the generated XUV. In an ideal case there is a rise in the gas density as the valve opens, so does the XUV yield grow, then it reaches a maximum, when the valve opens the most. The gas density corresponding to the maximum valve opening should stay fairly constant during the opening time of the valve, then it should slowly drop to zero with the closing valve. This is the typical behavior at the disk shaped piezo valve. We show that the gas density does not stay constant during the opening time of the cantilever piezo valve, which introduces an extra factor to optimize during the high harmonic generation. 

%%%%%%%%%%%%%%%%%%%%%%%%%%%%%%%%%%%%%%%%%%%%%%%%%%%%%%%%%%%%%%%%%%%%%%%%%%%%%%%%%%%%%%%%%%%%%%%%%%%%%%%%%%%%%%%%%%%%%
\begin{figure}[t!]
\begin{center}
    \includegraphics[width=0.48\textwidth]{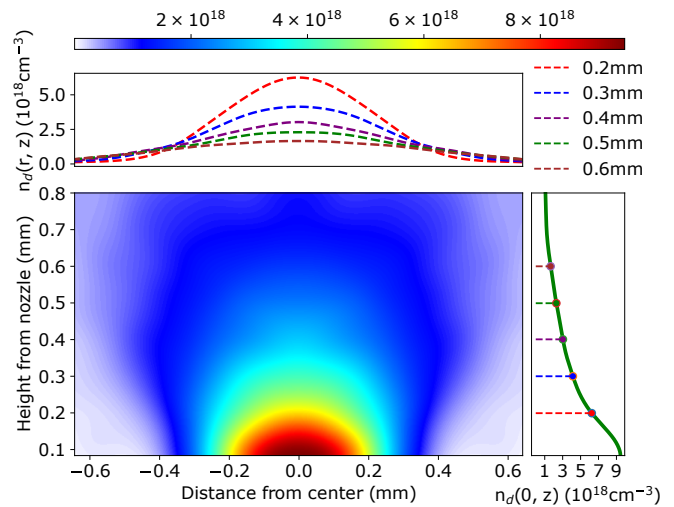}%
    \caption{\label{fig:single} (a) The distribution of the number density of the gas along the radial and the vertical direction. The corresponding colorbar is located at the top of the Figure and is expressed in atoms/cc. (b) The vertical decay of the gas number density from the center of the nozzle (radial position 0.0 mm) as one moves away from the nozzle (green line). The circles with their horizontal dashed lines show the vertical position of the outlines of the (c) radial gas density distribution. The actual vertical distances of these lineouts are showns in the top right corner.  }%
    \end{center}
\end{figure}
%%%%%%%%%%%%%%%%%%%%%%%%%%%%%%%%%%%%%%%%%%%%%%%%%%%%%%%%%%%%%%%%%%%%%%%%%%%%%%%%%%%%%%%%%%%%%%%%%%%%%%%%%%%%%%%%%%%%%%%

\section{\label{sec:results}Results}

\subsection{Experimental observations}

%%%%%%%%%%%%%%%%%%%%%%%%%%%%%%%%%%%%%%%%%%%%%%%%%%%%%%%%%%%%%%%%%%%%%%%%%%%%%%%%%%%%%%%%%%%%%%%%%%%%%%%%%%%%%%%%%%%%%%%%%%%%%%%%%%%%%%%%%%%%%%%
\begin{figure*}[th!]
    \includegraphics[width=0.97\textwidth]{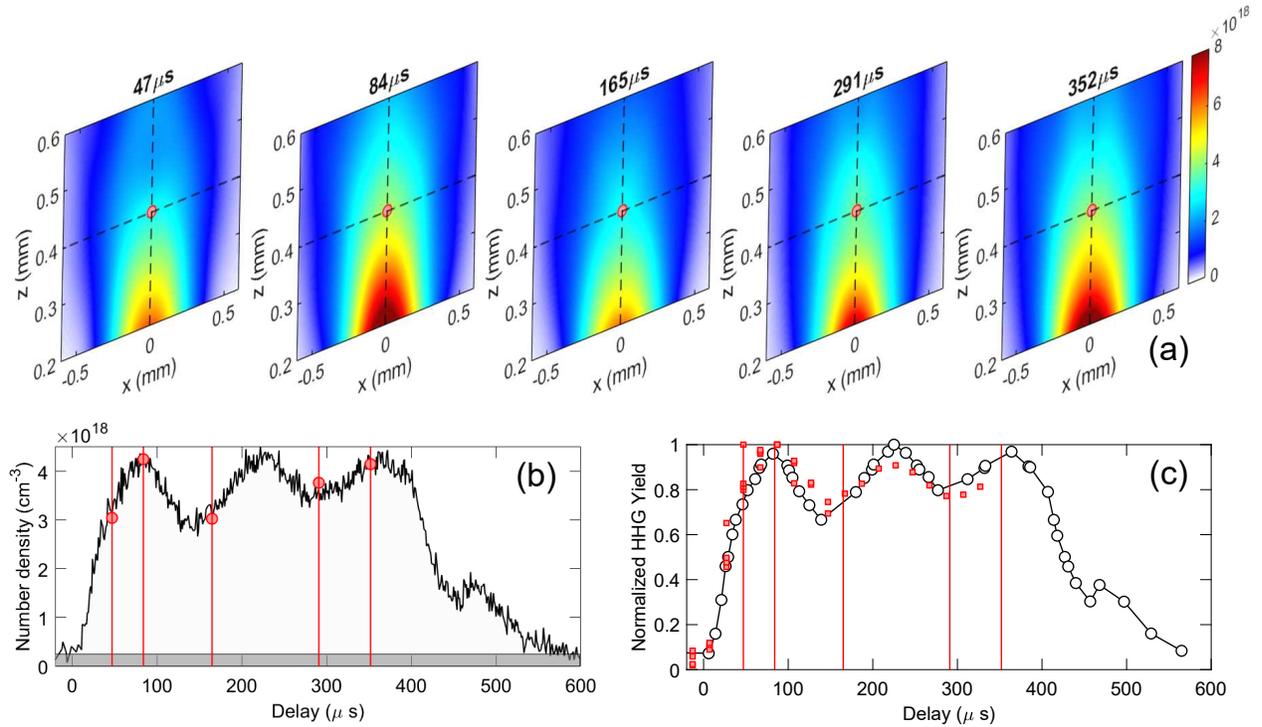}%
    \caption{\label{fig:slices} (a) The spatial distribution of measured 2D gas atomic density profiles $n_{a}(x,z,t)$ [$=n_{a}(y,z,t)$] as a function of delay from valve opening. The black dashed vertical line along $y$ defines the gas jet symmetry axis and the dashed horizontal line $x$ is along the laser propagation direction during strong field interaction. The colorbar represents atomic density in atoms/cc. The red shaded circle at the center of each gas density slice indicates the reference point where the interaction is centered. (b) Gas atomic density at the same point as a function of delay time (the black curve). The dynamics of the high repetition rate cantilever valve opening and closing leads to a characteristic variation of gas density as a function of delay. The vertical red lines correspond to the delays at which the spatial profile is presented in (a). The corresponding data points are indicated in red dashed circles in the figure. The grey shaded horizontal region indicates the detection threshold by our interferometric technique, below which data becomes noisy. (c) The red dashed rectangles depict the variation in high harmonic yield (normalized to the maximum) at different time delays. The black hollow circles demarcate the same, but this time calculated from simulations that mimic experimental gas density conditions. The variation in high harmonic yield in (c) shows striking resemblance to the gas density curve presented in (b).}%
\end{figure*}
%%%%%%%%%%%%%%%%%%%%%%%%%%%%%%%%%%%%%%%%%%%%%%%%%%%%%%%%%%%%%%%%%%%%%%%%%%%%%%%%%%%%%%%%%%%%%%%%%%%%%%%%%%%%%%%%%%%%%%%%%%%%%%%%%%%%%%%%%%%%%%%%%%
In Fig.~\ref{fig:single}(a) we present the retrieved gas atomic number density distribution along the radial (please note that the radial $r$, $x$ and $y$ distributions are same due to the cylindrical symmetry of the gas flow) and the vertical direction, achieved by applying the protocol presented in the previous section. The presented density map is achieved at one specific delay after opening of the gas valve. Due to the shape of the nozzle, the expanding gas jet is rather confined into a cylinder along the vertical direction with a diameter of 500 $\mu$m (which is the opening size of the nozzle) and not expanding much in the radial direction. As expected the maximum value of the number density distribution is close to the exit of the valve and its value is around $1.2\times10^{19}$ particles per cm$^3$. The central $z$ line-out $n_{d}(r=0,z)$ presented in Fig.~\ref{fig:single}(b) shows the exponential decay of the number density distribution as the distance from the nozzle increases in the vertical direction, a typical feature of such nozzle geometry \cite{Altucci1996,Sylla2012}. In Fig.~\ref{fig:single}(c), we  plot the radial line-outs of $n_{d}(r,z)$ for the five different z values marked in Fig.~\ref{fig:single}(b). 

In all the measurements, the opening window of the valve is set to 400 $\mu$s. In Fig.~\ref{fig:slices}\,(a) we present five different snapshots of the dynamic evolution of the gas number density distribution in space, within the opening window of the gas valve. As discussed before, the gas density is exponentially dropping as the function of height from the nozzle exit. Therefore, in order to access the higher density region for the HHG experiments, for the given focusing configuration, one has to shoot as close to the exit of the nozzle as possible, without damaging the nozzle. Because of technical constrains, for our conditions, in the HHG experiments, we kept the center of the intense focused beam around 400 $\mu$m above the exit. The black dashed horizontal lines on the colormaps in Fig.~\ref{fig:slices}\,(a) represent the laser propagation axis ($z=$400 $\mu$m) for HHG experiments. The red solid circles in Fig.~\ref{fig:slices}\,(a) indicate the position of maximum atomic gas density along the laser propagation axis. These colormaps show that during the opening time of the valve the laser focus experiences large variations in the gas distribution. In order to closely follow the temporal evolution we obtain a large number of 2D spatial gas number density snapshots as a function of delay within the opening window of the valve. In Fig.~\ref{fig:slices}\,(b) we plot the maximum number density seen by the center of the laser focal spot as a function of this delay. The vertical red lines in Fig.~\ref{fig:slices}\,(b) indicate the temporal delay where the 2D number density snapshots (presented in  Fig.~\ref{fig:slices}\,(a)) were taken while the red circles and the black curve correspond to maximum gas number density in the focal spot.

In an ideal case, during the scan of the opening window one would see the rise then the drop of the number of particles in the interaction region, while the maximum would show the right delay setting for optimal harmonic yield. However, in our case in Figure \ref{fig:slices}\,(b) we clearly identify several maxima as the function of the delay. Time dependent injection of the gas jet within the pulsed valve aperture time and a consequent gas density depletion has been observed by other research groups as well \cite{Altucci1996,Brandi2011,Hagmeister2022}. One can observe two important features on the graph in Fig.~\ref{fig:slices}\,(b): on one hand the opening and closing of the valve is not sudden, but takes several tens of microseconds. On the other hand, during the opening window, the signal oscillates. The first minimum is a drop of approximately 40 percent in the number density, while the following drops are less intense. However the maxima reaches roughly the same level in each case. The observed facts are a clear sign of a damped oscillation of the cantilever piezo \cite{Irimia2009} which is well known from vibrations of cantilever beams \cite{Repetto2012}. The frequency of this oscillation is roughly 7.6 kHz determined by physical parameters of the piezzo and not influenced by the driving frequency \cite{Irimia2009}. The results show that the dynamics of the cantilever when using it as a valve introduces variations in the gas jet density profile emphasizing the importance of such metrology in any experiment that is sensitive to the gas atomic number density. In addition, since the gas expansion from such a valve is non-trivial, the knowledge of the exact distribution of gas density could improve the understanding and modelling of the HHG process, while correlating with experimental observations.

Fig.~\ref{fig:slices}\,(c) presents the normalized high harmonic yield (red rectangles) as the function of the delay of the arrival time of the interacting intense focused laser pulse with respect to the opening time (marked as zero delay which signifies a measurable number density above the detection threshold in Fig.~\ref{fig:slices}(b)) of the valve. The relative high harmonic yield is experimentally measured with a thin film coated photodiode (Optodiode-AXUV100AL). The HHG yield data presented in Fig.~\ref{fig:slices}\,(c) is normalized with respect to the maximum measured yield. The red vertical lines correspond to the delay times presented in Fig.~\ref{fig:slices}\,(a). Multiple red rectangles at the same delay time (wherever available are presented in Fig.~\ref{fig:slices}\,(c)) represent typical fluctuations in the measured yield. The measured HHG yield data in Fig.~\ref{fig:slices}\,(c) follows remarkably well the number density variations presented in Fig.~\ref{fig:slices}\,(b).

For the HHG interaction, here we note the following points:
\begin{itemize}
    \item The laser pulse duration ($\sim$11$\,$fs FWHM of the pulse intensity envelop) is negligible on the timescale of gas density evolution. This implies that the interacting pulse sees a frozen gas density distribution in the transverse plane. This ensures that the microscopic emitter distribution across the focal spot, within the pulse duration during HHG are not evolving.
    \item The transit time of the intense laser pulse through the gasjet target (typically < 10$\,$ps in our case) is also negligible compared to the time scale of gas dynamics. This ensures that the measured temporal snap-shots of the spatial distribution of gas number density does not change during pulse propagation and thus can be utilised for macroscopic phase matching considerations in HHG.
    \item Since the confocal parameter ($\sim$49\,cm) is significantly larger than the medium length ($\sim$1\,mm) under our experimental configuration, we are not limited by longitudinal variation of intensity and the contribution of Gouy phase, associated with the spatial focusing of the fundamental laser pulse, to phase matching is unimportant.
    \item The gradient in the number density (presented in Fig.~\ref{fig:single}(b)), across the focal spot diameter of $\approx$300 $\mu$m can result in subtle effects like influencing the phase matching condition for the HHG. This can lead, for example, to distortions in the XUV wavefront impacting the focusability of the XUV beam \cite{Drescher2018,Hoflund2021}, which is beyond the scope of the present manuscript. 
\end{itemize}

The experimental results demonstrate that in case of HHG, it is essential to know the exact number density of the gas medium in a space time resolved manner. In addition, in order to optimize the harmonic yield one has to synchronize the arrival of the generating laser pulse with the opening time of the valve and introduce an appropriate relative time delay, depending on spatio-temporal the characteristics of the gas jet under utilization. 

At this point, we would like to emphasize that the monotonic nature of the HHG yield as a function of measured gas jet atomic density as observed experimentally and presented in Fig.~\ref{fig:slices}\,(c) is not the case in general. In case of coherent light emission - like HHG - the generated photon flux scales quadratically with the number of emitters under ideal conditions \cite{Heyl2016}. The resemblance between the jet density (Fig.~\ref{fig:slices}\,(b)) and the harmonic yield (Fig.~\ref{fig:slices}\,(c)) highlights the importance of phase matching, as it manifests under our specific experimental conditions. A close investigation of the correlation between the number density data in Fig.~\ref{fig:slices}\,(b) and the measurements in Fig.~\ref{fig:slices}\,(c) reveal that within our interaction regime the HHG yield is almost proportional to the gas pressure. Phase matching is a complex dynamical \cite{Tao2017, Fu2022} process and the relation between gas atomic number density and HHG yield is not straight forward in the short pulse regime. In order to investigate further we undertake numerical simulations in the following.

\subsection{Numerical validation using 3D Simulation}
Direct measurement of the gas number density distribution in the HHG interaction region is crucial not just from the optimization of the high harmonic source. Such metrology also enables one to feed experimental measurements into state of the art simulation tools that are often utilised to investigate the strong field interaction further. In this case the numerical simulations can be performed in a virtual experimental set up with initial parameters mimicking the real experimental conditions. This is important, if one needs to reconcile experimental observations with theoretical results and interpret the relevant physics in a correct manner. 

In our case, we undertake such an effort and use state of the art simulations where the gas jet metrology data is fed as input to simulate the harmonic yield. We note here, that the macroscopic effects like plasma generation, absorption and refraction during propagation play significant part in the phase matching process and hence cannot be neglected for calculation of the HHG yield. In order to investigate the experimental results further, we have performed a series of macroscopic simulations using a three-dimensional (3D) non-adiabatic model, described in detail elsewhere \cite{Tosa2009,Priori2000,Major2019}.\par
As a short summary, the simulation is performed in three self-consistent computational steps. Firstly, to analyze the propagation of the linearly polarised electric field of the fundamental laser pulse $E(\mathbf{r}_{L},t)$ in the generation volume, the nonlinear wave equationv of the form
\begin{equation}
\nabla^2E(\mathbf{r}_{L},t)-\frac{1}{c^2}\frac{\partial^2E(\mathbf{r}_{L},t)}{\partial t^2}=\frac{\omega_0^2}{c^2}(1-n_\mathrm{eff}^2(\mathbf{r}_{L},t))E(\mathbf{r}_{L},t) ,
\label{WaveEqn1}
\end{equation}
is solved \cite{Tosa2009}. In the previous equation $c$ is the speed of light in vacuum, $\omega_0$ is the central angular frequnecy of the laser field, and the suffix $L$ in $\mathbf{r}_{L}$ indicates that this vector represents the coordinate in the frame with respect to the laser axis (in contrast to the $r$ scalar coordinate described previously around the gas jet symmetry axis). The effective refractive index $n_\mathrm{eff}(\mathbf{r}_{L},t)$ of the excited medium --- depending on both space and time --- can be ontained by \cite{Tosa2016} $n_\mathrm{eff}(\mathbf{r}_{L},t)=n+\bar{n}_2 I(\mathbf{r}_{L},t)-\frac{\omega_p^2(\mathbf{r}_{L},t)}{2\omega_\mathrm{0}^2}$, where $I(\mathbf{r}_{L},t)=\frac{1}{2}\epsilon_0c\lvert\tilde{E}(\mathbf{r}_{L},t)\rvert^2$ is the intensity envelope of the laser field (note that in this expression the complex electric field $\tilde{E}(\mathbf{r}_{L},t)$ is present \cite{Chang2011}), and $\omega_p(\mathbf{r}_{L},t)=[n_e(\mathbf{r}_{L},t)e^2/(m\epsilon_0)]^{\frac{1}{2}}$ is the plasma frequency. The plasma frequnecy is well-known to be a function of the electron number density $n_e(\mathbf{r}_{L},t)$, and its expression also contains the electron charge $e$, the effective electron mass $m$, and the vacuum permittivity $\epsilon_0$). Dispersion and absorption, along with the Kerr effect, are thus incorporated via the linear ($n$) and nonlinear ($\bar{n}_2$) part of the refractive index. Absorption losses due to ionization \cite{Geissler1999} are also included, while plasma dispersion is estimated based on ionization values in the last term of $n_\mathrm{eff}(\mathbf{r}_{L},t)$. The model assumes cylindrical symmetry about the laser propagation direction $z_{L}$ ($\mathbf{r}_{L}\rightarrow r_{L},z_{L}$) and uses paraxial approximation \cite{Tosa2016}. Applying a moving frame translating at the the speed of light, and by eliminating the time derivative using Fourier transform $\mathscr{F}$, equation.~(\ref{WaveEqn1}) reduces to the explicit form,
\begin{multline}
\left(\frac{\partial^2}{\partial r_{L}^2}+\frac{1}{r_{L}}\frac{\partial}{\partial r_{L}}\right)E(r_{L},z_{L},\omega)-\frac{2i\omega}{c}\frac{\partial E(r_{L},z_{L},\omega)}{\partial z_{L}}\\=
\frac{\omega^2}{c^2}\mathscr{F}[(1-n_\mathrm{eff}^2(r_{L},z_{L},t))E(r_{L},z_{L},t)].
\label{WaveEqn1r}
\end{multline}
Equation.~(\ref{WaveEqn1r}) is solved using the Crank–Nicolson method in an iterative algorithm \cite{Tosa2016}. The ABCD-Hankel transform  is used to define the laser field distribution in the input plane of the medium \cite{Ibnchaikh2001, Major2018AO}.\par

In step two, we calculate the single-atom response (dipole moment $D(t)$) based on the laser-pulse temporal shapes available on the complete ($r_{L}, z_{L}$) grid, by evaluating the Lewenstein integral \cite{Lewenstein1994,Nayak2019}. The macroscopic nonlinear response $P_{nl}(t)$, is then calculated by taking the depletion of the ground state into account \cite{Lewenstein1994,Le2009} using $P_{nl}(t)=n_aD(t)\mathrm{exp}\bigg[-\int\limits^{t}_{-\infty}w(t')dt'\bigg]$, where $w(t)$ is the ionization rate obtained from tabulated values calculated using the hybrid anti-symmetrized coupled channels approach (haCC) \cite{Majety2015} showing a good agreement with the Ammosov-Delone-Krainov (ADK) model \cite{ADK1986} and $n_a$ is the atomic number density within the specific grid point ($r_{L}, z_{L}$) \cite{Tosa2016,Gaarde2008}.\par

In the third step we calculate the propagation of the generated harmonic field $E_h(\mathbf{r}_{L},t)$ using the wave equation
\begin{equation}
\nabla^2E_h(\mathbf{r}_{L},t)-\frac{1}{c^2}\frac{\partial^2E_h(\mathbf{r}_{L},t)}{\partial t^2}=\mu_0\frac{d^2P_{nl}(t)}{dt^2} ,
\label{WaveEqn2}
\end{equation}
with $\mu_0$ being the vacuum permeability. Equation.~(\ref{WaveEqn2}) is solved in a manner similar to equation.~(\ref{WaveEqn1}), but without an iterative scheme (since the source term is known). The amplitude decrease and phase shift of the harmonic field - caused by absorption and dispersion, respectively - are incorporated at each step when solving equation~(\ref{WaveEqn2}) by taking into account the effect of complex refractive index on wave propagation. The real and imaginary parts of the refractive index in the XUV regime are from tabulated values of atomic scattering factors \cite{Henke1993}.\par

The simulation method described above assumes radial symmetry aroung the laser propagation axis. For the laser spatio temporal profile we use the measured focal spot distribution and experimental laser pulse duration in order to mimic the real experimental conditions. For the gas jet atomic number density profile we use the measured gas jet number density profile along the axis of laser propagation (peak densities as shown in Fig.~\ref{fig:slices}\,(b)). Thus, within our numerical simulations, the influence of gas density gradient across the laser focal spot (along the symmetry axis of the gas jet as presented in Fig.~\ref{fig:single}(b)) is lumped into an average value.

%%%%%%%%%%%%%%%%%%%%%%%%%%%%%%%%%

Figure \ref{fig:slices}\,(c) presents tbe simulated harmonic yield (black hollow circles) as the function of the delay from the opening of the valve. The gas jet pressure for the simulation was calculated from the number density variation on Figure \ref{fig:slices}\,(b). Both the HHG measurements and simulations show a remarkable resemblance to the jet density variation measured with the interferometric technique.

The simulations also revelaed that under the circumstances that describe these experiments, transient phase matching \cite{Schotz2020, Schutte2015} limits efficient generation to the first half of the short laser pulse. At the same time, due to minimal reshaping of the pulsed laser beam, there are spatially homegenous phase matching conditions in the whole interaction volume. This allows us to apply a simple model\cite{Constant1999} to explain the variation of the observable harmonic flux in the absorbing medium. The analysis confirmed that with the coherence lengths and absorption lengths involved, the harmonic flux changes close to linearly with the change of atom number density.

\section{\label{sec:conclusion}Conclusion}
On one hand, an interferometric gas density characterization was developed for underdense gas jets produced from a high frequency (up to 5 kHz) cantilever piezo valve. On the other hand we show that the cantilever valve has its characteristic dynamics while opening the valve resulting in the oscillation of the gas density as the function of time. Using HHG from such a gas jet target, we observe a remarkable experimental correlation in between the gas density and HHG yield. Our results have been corroborated by sophisticated simulations that self consistently include both microscopic HHG and macroscopic propagation effects under conditions mimicking the real experimental scenario. Our results establish the feasibility of utilizing cantilever based high repetition rate gas valves for high harmonic generation processes, emphasizing the importance of precise timing control in order to access proper gas density regime. This also shows that appropriate time and space resolved characterization and monitoring of such gas valves is an important aspect for its application and reproducible performance is easily achieved by properly managing the synchronization of gas jet with respect to the arrival time of the laser. The results are also important to a diverse field of studies which can benefit from high repetition rate gas jets, where the signature effects of the phenomena, have sensitive dependence upon the precise gas density profile such as molecular or atomic quantum path interferometry \cite{ Chatziathanasiou2019,Csizmadia2021}, ion spectroscopy from dilute plasma \cite{ Lifschitz2014,Malka2016} spatio-temporal \cite{ Wikmark2019} and equivalently spatio-spectral \cite{ Dubrouil2014,Hoflund2021} control of attosecond pulses, and in designing of gas based extreme-ultraviolet refractive optics \cite{Drescher2018}, to name a few.

\section{\label{sec:acknowledgments}Acknowledgments}
ELI ALPS is supported by the European Union and co-financed by the European Regional Development Fund (ERDF) (GINOP-2.3.6-15-2015-00001). This project has received funding from the European Union Framework Programme for Research and Innovation Horizon 2020 under IMPULSE grant agreement No 871161. S.K. acknowledges Project No. 2020-1.2.4-TÉT-IPARI-2021-00018, which has been implemented with support provided by the National Research, Development and Innovation Office of Hungary, and financed under the 2020-1.2.4-TET-IPARI-CN funding scheme. 

\bibliography{ref.bib}

\end{document}